# The Astronomical, Astrobiological and Planetary Science Case for Interstellar Spaceflight

I. A. Crawford

Department of Earth and Planetary Sciences, Birkbeck College London
Malet Street, London, WC1E 7HX

**Abstract**

A review is presented of the scientific benefits of rapid (v ≥ 0.1$c$) interstellar spaceflight. Significant benefits are identified in the fields of interstellar medium studies, stellar astrophysics, planetary science and astrobiology. In the latter three areas the benefits would be considerably enhanced if the interstellar vehicle is able to decelerate from its interstellar cruise velocity to rest relative to the target system. Although this will greatly complicate the mission architecture, and extend the overall travel time, the scientific benefits are such that this option should be considered seriously in future studies.

## 1. Introduction

There can be no doubt that science, especially in the fields of astronomy, planetary science and astrobiology, will be a major beneficiary of the development of rapid (v ≥ 0.1$c$; where $c$ is the speed of light) interstellar spaceflight, as envisaged by the Daedalus [1] and Icarus [2] projects. In its long history astronomy has made tremendous advances through studying the light that reaches us from the cosmos, but there is a limit to the amount of information that can be squeezed out of the analysis of starlight and other cosmic radiation. Already we can identify areas where additional knowledge will only be gained by making *in situ* observations of distant astronomical objects. As noted in an earlier review of interstellar spaceflight [3], a sense of the scientific potential may be glimpsed by considering "the advantages of taking thermometers, magnetometers, mass-spectrometers, gravimeters, seismometers, microscopes, and all the other paraphernalia of experimental science, to objects that can today only be observed telescopically."

As reviewed by Webb [4] in the context of the *Daedalus* study, the scientific case for interstellar spaceflight naturally breaks down into four main areas: (i) studies of the interstellar medium conducted *en route* to a target star (possibly augmented by other astronomical investigations able to make use of the vehicle as an observing platform during the cruise phase); (ii) astrophysical studies of the target star itself (or stars if a multiple system is selected); (iii) studies of the target star's planetary system (if present); and (iv) biological studies of any life forms which may have evolved within this planetary system. Each of these areas has different requirements for the overall architecture of an interstellar mission, and for the scientific payload to be carried. Because the aim of this paper is to help identify options for *Icarus*, and other proposals for interstellar missions which may be made in the future, we here address the scientific opportunities of interstellar spaceflight without being constrained *a priori* to a particular mission architecture (as was the case for the *Daedalus* scientific payload [4]; see also the accompanying paper by Long et al. [2]).

## 2. Interstellar studies

By definition, any interstellar vehicle will have to traverse the interstellar medium between the Solar System and the target star. As any target star for an early interstellar mission is certain to be within a few parsecs (pc) of the Sun, and would very likely be the α Centauri (α Cen) system at a distance of 1.35 pc (4.40 light-years [5]), it follows that only the local interstellar medium (LISM) is relevant here (a discussion of other potential target stars, all also within the

LISM, is given by [2]). In the years since the original Daedalus study our knowledge of the LISM has improved significantly (see, e.g., [6-9]). It is generally accepted that the Sun is currently located close to the boundary of a small (spatial extent ≤ 3 pc); low density ($n_H$ ~ 0.1-0.2 $cm^{-3}$, where $n_H$ is the density of hydrogen nuclei), warm (T ~ 7500 K), partially ionised interstellar cloud known as the Local Interstellar Cloud (LIC). Whether the Sun lies just within, or just outside, the LIC is currently a matter of debate [8,9]. The LIC is only one of a several broadly similar interstellar clouds within a few pc of the Sun – Redfield & Linsky [8] identify seven within 5 pc. These are immersed in the very empty ($n_H$ ~ 0.005 $cm^{-3}$) and probably hot (T ~ $10^6$ K) Local Bubble (LB) in the interstellar medium, which extends for about 60-100 parsecs from the Sun in the galactic plane before denser interstellar clouds are encountered (at high galactic latitudes the LB appears to be open, forming a chimney-like structure in the interstellar medium which extends into the galactic halo; e.g. [10] and references cited therein).

The properties of the LIC, and other nearby clouds, have been determined by spectroscopic studies of interstellar absorption lines towards nearby stars, augmented in the case of the LIC by observations of interstellar matter entering the Solar System and interacting with the heliosphere [9]. Direct measurements of the interstellar material immediately beyond the heliopause (which likely forms the boundary of the LIC [8,9]) will probably be made by precursor interstellar space probes operating to distances from a few hundred to perhaps a thousand AU (i.e. ≤0.005 parsec) within the next half century or so (e.g. [11-15]). However, direct measurements of more distant interstellar material, including the properties of nearby clouds other than the LIC, will have to await the development of a true interstellar probe such as that under consideration here.

Key measurements that could be made from an interstellar vehicle, and which would add enormously to our understanding of interstellar processes, would include *in situ* determinations of density, temperature, gas-phase composition, ionisation state, dust density and composition, interstellar radiation field and magnetic field strength, all as a function of distance between the Sun and the target star system. Assuming measurements of these properties could be made once per day then, for a spacecraft travelling at $0.1c$, the sampling interval would be 17 AU (i.e. about half the radius of the Solar System), which would yield unprecedented knowledge of the structure of the LISM not obtainable in any other way (as astronomical observations of the interstellar medium only determine properties averaged over the whole sightline, and then often with large uncertainties). Of course, even higher spatial resolution measurements may be possible, and would be especially valuable while traversing the heliosphere, and the corresponding astrosphere of the target star.

Given that the α Cen system will very likely be the target of the first interstellar mission this sightline deserves special attention. Absorption line measurements indicate that α Cen itself lies beyond the LIC and that the line-of-sight is dominated by another nearby interstellar cloud – dubbed the 'G' cloud because it lies in the Galactic centre direction [7,8] (a sketch showing the relative positions of the Sun, LIC and G cloud is given in Fig. 1 of Lallement *et al.* [7]). The properties of the G cloud are broadly similar to those of the LIC, although it appears to have a somewhat lower temperature ($T \sim 5500$ K) and possibly a lower depletion of heavy elements [8]. It is currently uncertain whether the Sun is embedded in the LIC, or lies just outside it in a region where the LIC is interacting with the G cloud [8,9]. An interstellar mission to α Cen would resolve this matter, if it is not resolved earlier by interstellar precursor missions. If the Sun does lie within the LIC, then a mission to α Cen would sample the outer layers of the LIC, an interval of low density LB material, the edge of the

G cloud, and the deep interior of the G cloud. This would sample one of the most diverse ranges of interstellar conditions of any mission to another star located with 5 pc of the Sun, as most other potential targets lie within the LIC (see Table 1 of Redfield and Linsky [8]). Even if the Sun lies just outside the LIC (as argued by Redfield and Linsky [8]), the trajectory to α Cen would still permit detailed observations of the boundary of the G cloud (and its possible interaction with the LIC), and determine how its properties change with increasing depth into the cloud from the boundary.

Such precise *in situ* measurements of interstellar material, even though obtained on a very local scale in the galactic context, would be invaluable for validating ('ground truthing') inferences based on astronomical techniques which will, of necessity, continue to be used to determine interstellar medium properties at larger distances, both within our Galaxy and beyond. These measurements will also be invaluable for the planning of all future interstellar space missions. The first mission will be a pathfinder in this respect, and will enable all subsequent missions to be designed with a much firmer knowledge of the properties of the material through which they will have to travel. Direct measurements of the interstellar dust density, and the *size distribution* of dust particles, will be especially important because sub-micron sized dust particles will erode exposed surfaces (see the analysis performed for *Daedalus* by Martin [16]), and larger particles (i.e. larger than a few tens of microns) may, if present in the LISM, cause a catastrophic failure of the vehicle. Determining some of the other properties of the LISM will also be important for longer term planning of other interstellar propulsion concepts – for example, determining the LISM hydrogen density, and its ionisation state, will be important for assessing the future practicality of interstellar ramjets ([17]; see also the brief review in [3] and references therein).

Before leaving science which may be performed during the cruise phase, we should also mention the possible use of the very long baseline for trigonometrical distance determinations of distant galactic and extra-galactic objects. This was highlighted as a key cruise phase science objective for *Daedalus* [4]. However, since the *Daedalus* study there have been dramatic improvements in the capabilities of space-based astrometry. Future instruments such as the European Space Agency's *Gaia* mission [18] and NASA's Space Interferometry Mission [19] will probably be able to measure parallaxes to individual stars throughout the Galaxy and beyond within the next decade. Moreover, by making use of the baseline produced by the Sun's motion through the Galaxy (~800 AU over ten years) these instruments may even be able to obtain parallaxes of quasars at giga-parsec distances [20]. Given the advances in this field that are likely long before rapid interstellar travel becomes a possibility, it is no longer clear how much value the ~1 pc baselines enabled by the latter will be for observational astrometry. Nevertheless, it is still worth keeping an open mind to the possibilities for cruise phase astronomical observations during an interstellar mission, as these could presumably be implemented at very little marginal cost.

Finally, we note that none of the measurements discussed in this section impose stringent constraints on the architecture of an interstellar mission. From the perspective of interstellar medium and astronomical studies a simple undecelerated sub-relativistic (v $\geq$ 0.1$c$) flyby mission would be sufficient, although steps would have to be taken to protect the instruments from damage induced by high speed collisions with the dust component of the interstellar medium.

## 3. Stellar studies

We know far more about the Sun than any other star, simply by virtue of the fact that it is so close to us. Interstellar spaceflight would enable us to obtain comparable information about stars of other spectral types (or, in the particular case of α Cen A, the same spectral type but of a different age and metallicity). Such observations are likely to lead to significant advances in stellar astrophysics, although their extent will depend, at least in part, on the time window which is available to make them, and this will have implications for the mission architecture.

There are really three reasons for our enhanced knowledge of the Sun compared to other stars: (i) vastly increased spatial resolution, which permits the observation of small scale features on the photosphere (e.g. sunspots and associated phenomena), chromosphere and corona; (ii) greatly increased brightness, which permits very high-signal-to-noise observations (which, among other things, facilitates the use of helioseismology to probe the Sun's interior structure); and (iii) a long time base of observations (hundreds of years of recorded human observations, and millions of years of relevant geological records on the Earth and other planets).

Although we might expect interstellar space travel to help principally with the first two of these, we have to recognize that, long before rapid interstellar spaceflight becomes feasible, astronomical instrumentation is likely to have advanced to the point where many nearby stars will be resolvable from observations conducted from the Solar System. Indeed, we have already reached the point where the radii of nearby low-mass stars can be measured directly using ground-based optical interferometry [21], and the next-generation of space-based interferometers may be able to resolve surface features [22] (which is already possible for giant stars [23]). Similarly, the advent of very large ground and space-based telescopes will go some way to address signal-to-noise limitations caused by the relative faintness of other stars compared to the Sun. Nevertheless, it will always be true that the spatial resolution and signal-to-noise of observations conducted from a vantage point within a few AU of a target star will always be higher than comparable observations attempted from the vicinity of the Earth. Thus, while we should avoid exaggerating the benefits to observational stellar astronomy from interstellar missions to the closest stars, we can nevertheless be sure that such advantages do exist.

While undoubtedly scientifically valuable, stellar observations conducted from an interstellar flyby mission would suffer from the same disadvantages, arising from the short time span available for the highest resolution observations, as would the planetary science studies described in Section 4 below. Much greater advantages would result if it proved possible to decelerate at the target star system. It would then be possible to ring the star with satellites to acquire long term observations of the whole stellar surface and to obtain time-resolved, high-resolution, multi-wavelength observations of the corona and solar wind for comparison with the Sun. Examples of the kinds of observations which would then be possible are provided by the SOHO and STEREO missions [24,25]. As observations from Solar satellites such as these are of demonstrable importance for understanding of the Sun, it follows that they would also be desirable for studies of other stars, but they will require the interstellar carrier spacecraft to decelerate essentially to rest in the target star system.

A mission architecture permitting deceleration to rest in the target system would also make it possible to examine *geological* records of longer term stellar activity preserved on any planets that may be present (see Section 4). In particular, it might permit the derivation of an independent age for the stellar system through *in situ* radiometric dating of primitive asteroidal materials (if present and identifiable), which would be valuable as a check on stellar age derived by the astronomical method of isochron fitting in the Hertzsprung-Russell diagram. Clearly, the better these methods are calibrated for stars of different spectral types, the better will be our age estimates for more distant objects for which *in situ* measurements may never be possible.

Finally, we note that all stars are surrounded by circumstellar matter to varying degrees, and *in situ* studies of this would also be of scientific interest. Undoubtedly of greatest interest would be studies of protoplanetary disks from which planets may have recently formed, or still be forming. *In situ* observations of the density, temperature, magnetic field and, crucially, dust particle size, as a function of radial distance from the star and distance from the disk mid-plane would greatly add to our understanding of planet formation processes. However, the nearest known example of a circumstellar disk of this type is around the star epsilon Eridani (ε Eri) at a distance of 3.2 pc [5,26] and, although a possible candidate for an early interstellar mission, its relatively large distance means it is unlikely to be a high priority for the first such mission.

All things considered, from a stellar astrophysics viewpoint, the α Cen system appears the most attractive of all the nearby possible targets, as it would permit close-up observations of stars of at least two different spectral types (G2V and K1V for α Cen A and B, respectively). Moreover, it may be a relatively straightforward to launch a sub-probe from the main vehicle after its main acceleration phase to flyby Proxima Centauri (separated by only 2.18 degrees from α Cen A/B on the sky), which would permit our first close up view of a red dwarf star (spectral type M6V).

## 4. Planetary Science

Over 400 planets are now known to orbit other stars [27], with new discoveries being made every month. A conservative view of the statistics to-date implies that at least 5% of solar-type stars have planets [28]. However, the known biases in present detection methods imply that many more planets exist than can be detected, and allowance for this raises the estimate to ~ 9% of solar-type stars having planets with masses greater than ~0.3 Jupiter masses and orbital periods less than 13 years [28]. Other estimates of the fraction of stars with planets are considerably higher. In particular, improvements in detection sensitivity have led to the identification of lower mass planets, and Mayor *et al.* [29] estimate that 30% of solar-type stars may have planets with masses less than 30 Earth masses (including many so-called 'super Earths' with masses between 1 and 10 Earth masses). Given that Earth-mass planets themselves (as well as giant planets with orbital periods much longer than that of Jupiter) are not yet detectable, it is entirely possible that most stars have planets. However, even when planets are detected, for the most part current techniques can do little more than determine lower limits to their masses and establish their basic orbital parameters (period, semi-major axis, and eccentricity). In the relatively rare cases of transiting planets (approximately 15% of the known exoplanets [27]) it is also possible to determine the size (and thus density) of exoplanets, and in some cases obtain rudimentary spectral information on the composition of their upper atmospheres (e.g. [30]).

Future astronomical observations are certain to improve on our knowledge of planetary systems around nearby stars. Indeed the *Kepler* satellite should provide a reliable statistical estimate of the abundance of rocky terrestrial planets around other stars within the next few years [31]. These discoveries are likely to be followed in the coming decades by observations conducted with increasingly sophisticated space-based telescopes, such as the proposed *Darwin* instrument [32], able to directly image planets around stars within about 10 pc of the Sun and to obtain spectroscopic measurements of their atmospheres. Some longer-term possibilities are discussed by Schneider *et al.* [33] (although the brief discussion of direct investigation of exoplanetary systems by interstellar spacecraft in the latter part of this paper would have benefited from a more thorough review of the relevant literature).

It is salutary to reflect that, within the coming decades, astronomical observations will very likely have raised our knowledge of planetary systems around nearby stars to a level comparable to that obtained for the planets in our own Solar System prior to the space age. That is to say, we will know the number of planets in each system (down to some minimum mass that will probably be significantly less than that of Earth), together with their orbital parameters, masses and densities, presence or absence of an atmosphere, atmospheric composition, presence of large natural satellites, etc. All this can probably be learned without leaving the Solar System. However, the history of the exploration of the Solar System shows that obtaining significantly more knowledge of extrasolar planetary systems will require *in situ* observations by spacecraft. We can be sure of this because, over the last half century, spacecraft have completely revolutionised the study of the planets of the Solar System, providing information that could never have been obtained telescopically from the surface of the Earth or its immediate vicinity. To highlight just three out of hundreds of possible examples, consider the structure of the lunar interior as probed by the Apollo seismic experiments, the fine scale (i.e. mm to cm) resolution of mineralogical and sedimentary structures at the landing sites of the Mars Exploration Rovers (with their implications for the volcanic and hydrological histories of that planet), and the discovery of lakes of liquid methane (and indeed an entire methane hydrological cycle) under the orange smog of Titan's atmosphere by the Cassini/Huygens mission. It follows that if we wish to obtain comparable knowledge of the planets orbiting other stars then we will have to go there and look.

The analogy with the exploration of our own Solar System has implications for the architecture of an interstellar mission designed with planetary science in mind. There is a hierarchy of architectural options for planetary missions, in order of increasing complexity and energy requirements, but also in increasing scientific return: (i) fly-by missions; (ii) orbital missions; (iii) hard landers (including penetrators [34]); (iv) soft landers (with or without rover-facilitated mobility); and (v) sample return. The same general ordering will apply in the study of extrasolar planetary systems, although the relative jumps in difficulty between them are not the same in the two cases, as described below.

An undecelerated flyby will be the easiest to implement, and for this reason was adopted in the *Daedalus* study. However, the exploration of the Solar System shows that, while appropriate for the initial reconnaissance of a planetary body, flybys are very limited in terms of the knowledge they are able to collect (and sometimes this information can be misleading, as in the case of the Mariner 4 flyby of Mars in 1965 which revealed a lunar-like landscape and gave little intimation of the geological diversity discovered by later missions). The limitations of fly-bys in an interstellar mission will be exacerbated by the high speeds involved – the *Daedalus* study proposed to conduct planetary investigations from multiple sub-probes flying close to target planets at 12% of the speed of light [1]. This would permit less than a second of time available for detailed observations at distances comparable to the radii of planetary-sized bodies, although perhaps several hours of useful observations might be obtained on the approach to, and departure from, the planet in question.

Much more scientific information would be obtained if it proved possible to decelerate an interstellar vehicle (or at least any sub-probes designed to conduct planetary observations) from its interstellar cruise velocity. The benefits will be immediately obvious by comparing the results of the initial fly-by reconnaissance of Mars by Mariners 4, 6 and 7 with those of the early orbital missions (i.e. Mariner 9 and Vikings 1 and 2) which discovered, amongst other things, the giant Tharsis volcanoes, the Valles Marineris canyon system, and numerous dried up river valleys indicating a warmer, wetter Martian past. Of course, even more detailed information has resulted from the handful of soft landers and rovers that have successfully reached the surface.

Although in terms of Solar System exploration there is a big jump in energy requirements between orbital missions and soft (or even hard) landers, this would not be a major consideration in terms of an interstellar mission -- the energy differential between orbital insertion and a soft landing is trivial in comparison to that of decelerating a probe from a significant fraction of the speed of light. As for Solar System missions, landers would permit a range of geochemical, geophysical and astrobiological investigations that are simply not possible from an orbiting spacecraft. Thus, despite the added complexity involved, the potential scientific benefits are such that the designers of any interstellar mission capable of decelerating at its destination should consider including sub-probes that are capable of landing on the surfaces of suitable planets. This would be in addition to providing planetary orbiters (which will in any case be needed as communication relays if landers are deployed).

The most ambitious Solar System missions involve sample return, which allow detailed investigation of planetary materials in terrestrial laboratories. For any reasonable extrapolation of foreseeable technology, sample return is essentially impossible from an extrasolar planetary system on any reasonable timescale, as it would require no less than four separate sub-relativistic (≥0.1c) Δv increments. Nevertheless, in the context of a mission architecture where the main interstellar vehicle itself comes to rest in the target planetary system, it is possible to envisage sub-probes capable of landing on a planetary surface and returning samples to the main vehicle for more detailed analyses than would be possible on the sub-probe itself. While undoubtedly a complication, this would greatly enhance the planetary science return of an interstellar mission, and would also be valuable for many astrobiology investigations (as discussed in Section 5).

Before leaving this section we note that the nearest star for which there is currently reasonably secure evidence for a planetary system is ε Eri at a distance of 3.2 pc [35,36]; this is also the closest star with a known circumstellar disk [26]. There are eight known stars (or stellar systems) closer than ε Eri that would be easier targets for a first interstellar mission. Knowledge of whether any of them have planets, and are thus of comparable interest from a planetological (or astrobiological) perspective must await future discoveries. Of course, the α Cen system is of particular interest in this regard – no planets have yet been detected and there is an ongoing debate as to the extent to which the binary nature of the α Cen A/B system may have impeded the formation of planets (e.g. [37-39]). On the other hand, the system also contains the red dwarf Proxima Centauri, and most of the other stars closer than ε Eri are also red dwarfs, so recent results indicating that planets may be common around such stars [40] augur well for the likelihood of planetary systems closer than ε Eri. Only further observations will tell.

## 5. Astrobiology/Exobiology

Astrobiology is the science relating to the search for life elsewhere in the Universe, and especially the astronomical and planetary environments which may nurture it. By adding to our knowledge of other stellar and planetary environments, the *in situ* scientific investigations outlined above would be of considerable astrobiological value even if no indigenous life is present in the target system (just as studies of lifeless bodies in our own Solar System are of astrobiological relevance; e.g. [41,42]). Nevertheless, it is clear that the greatest scientific interest would be in the discovery and characterisation of any life forms which may be present. If such extraterrestrial organisms are found, their study will presumably become the subject of a new sub-discipline of biology where, by definition, the study of living things properly belongs [43].

As noted in Section 4, before rapid interstellar space travel becomes possible advances in astronomical techniques will probably have already identified which of the nearest stars are accompanied by planetary systems. Indeed, we are likely to know the basic architecture of these systems in some detail, and Solar System-based instruments will have the capability of detecting any molecular biosignatures that may be present in the atmospheres and/or on the surfaces of these planets [32,33]. Of course, the absence of a detectable biosignature does not necessarily mean that life is absent (an instrument such as *Darwin* [32] may not have found any evidence for life on Earth prior to the build up of oxygen in the atmosphere about 2.3 billion years ago, yet life was certainly present much earlier [44]). That said, we can be reasonably confident that astronomical observations will be able to establish a hierarchy of priorities among any planets which may be detected around the nearest stars: (i) planets where *bona fide* biosignatures are detected; (ii) planets that appear habitable (e.g. for which there is spectral evidence for water and carbon dioxide, but no explicit evidence of life being present); and (iii) planets which appear to have uninhabitable surfaces (either because of atmospheric compositions deemed non-conducive to life or because they lack a detectable atmosphere), but which might nevertheless support a subsurface biosphere. Thus, when planning an interstellar mission with astrobiology/exobiology in mind, we are likely to have a priority list of target systems prepared well in advance.

As for the planetary science cases discussed in Section 4, and for the same reasons, it is not immediately obvious that simple flyby missions could add significantly to information likely to be obtained by the astronomical techniques available at the time. There will be some advantages: for example, even travelling at 0.1c, sub-probes targeted to fly close to planets could presumably perform much more detailed analyses of their atmospheric compositions, especially trace constituents, than would be possible astronomically from the Earth. Nevertheless, it seems clear that only an interstellar probe that decelerated into its target star system would be able to deploy the kind of instrumentation that biologists would need to begin an investigation of an alien biosphere in any detail.

We can get an idea of the kind of instruments that would be required by considering those that have either been used (e.g. the *Viking* biology package [45] and the *Phoenix* high-resolution microscope [46]), or are planned to be used (e.g. the *Urey* organic molecule analyser [47] and the Life Marker Chip [48]), in the search for life on Mars. Doubtless more sophisticated biological tools will be available at the time of the first interstellar mission. However, it seems clear that deployment of instruments such as these would require the soft-landing of suitably instrumented sub-probes on a planetary surface. More detailed biological analyses may require the sub-probes to collect samples from the planetary surface and transport them to the main vehicle for more detailed analyses. None of this can be done flying through the target system at ten percent of the speed of light.

## 6. Conclusions

The principal conclusions of this paper are as follows:

(1) Considerable scientific advantages will result from the development of a fast (v ≥ $0.1c$) interstellar spaceflight capability, especially in the fields of interstellar medium studies, stellar astrophysics, planetary science and astrobiology.

(2) Important new knowledge of the structure and physical state of the local interstellar medium could be obtained from an interstellar mission sent to any of the stars within a few pc of the Sun. The direction to α Cen was shown to be of particular interest. In addition to their scientific importance, these measurements would help define the properties of the local interstellar medium for all later interstellar space missions. They pose few constraints on mission architecture, and could be conducted during the cruise phase of an undecelerated interstellar flyby.

(3) In the areas of stellar astrophysics, planetary science and astrobiology the scientific benefits would be considerably enhanced if the interstellar vehicle were able to decelerate from its interstellar cruise velocity to rest relative to the target system. Indeed, without this capability, an interstellar mission may not be able to add significantly to knowledge that is likely to be obtainable by Solar System-based astronomical instruments in the same timeframe. Thus, despite the added complications to the mission architecture, and the increased mission duration, it is strongly recommend that a deceleration capability be seriously considered for the *Icarus* study [2] and other plans for future interstellar space missions.

(4) The relative proximity of α Cen, together with its interesting interstellar sightline and the presence of stars of three different spectral types, makes it an attractive target for humanity's first interstellar mission. However, as the bulk of the scientific benefits of interstellar spaceflight pertain to planetary science and astrobiology, a final prioritization must await future developments in the detection of planetary systems around the nearest stars. Fortunately, expected advances in astronomical instrumentation over the next century should ensure that a comprehensive list of prioritized targets is available well before rapid interstellar travel is technically feasible.

(5) In the particular case of α Cen, if planets are discovered around either (or both) components A and B the ideal architecture for an interstellar mission would be one which decelerates into the A/B system, but which also launches an undecelerated flyby probe to Proxima Cen (located 2.18 degrees away). On the other hand, should Proxima Cen be discovered to harbour a planetary system, and α Cen A/B not, then it may be appropriate to decelerate at the Proxima system and send a flyby probe to α Cen A/B. The practicalities of such a mission architecture should be considered in future studies.

**Acknowledgements**

I would like to thank Kelvin Long for his help in organising the BIS symposium '*Project Daedalus - Three Decades On'*, at which this paper was presented. I would also like to record my thanks to the members of the original *Daedalus* Team for their pioneering study [1] which has been an inspiration behind much of the technical and popular literature on interstellar space travel for the last three decades.